\documentclass[fleqn,twoside]{article}
\usepackage{espcrc2}

\usepackage{graphicx}
\usepackage{epsfig}
\usepackage[figuresright]{rotating}

\newcommand{\AmS}{{\protect\the\textfont2
  A\kern-.1667em\lower.5ex\hbox{M}\kern-.125emS}}

\title{A Novel Approach to Non linear Shock Acceleration}

\author{P. Blasi\address[arcetri]{Osservatorio Astrofisico di Arcetri\\ 
        Largo Enrico Fermi, 5 - 50125 Firenze, ITALY}
        \thanks{blasi@arcetri.astro.it}}
       
\begin{document}

\begin{abstract}
First order Fermi acceleration at astrophysical shocks is often invoked
as a mechanism for the generation of non-thermal particles. This 
mechanism is especially simple in the approximation that the accelerated
particles behave like test particles, not affecting the shocked fluid. 
Many complications enter the calculations when the accelerated particles
have a backreaction on the fluid, in which case we may enter the non linear
regime of shock acceleration. In this paper we summarize the main features of
a semi-analytical approach to the study of the non linearity in shock
acceleration, and compare some of the results with previous attempts and
with the output of numerical simulations.
\vspace{1pc}
\end{abstract}

\maketitle

\section{Introduction}
Shock acceleration has been studied carefully and a vast literature
exists on the topic, including some recent excellent reviews 
\cite{be87,malkov}. Hence we do not try to provide here
any extensive introduction to the problem, but we rather limit ourselves to 
summarize some of the open issues related to the backreaction of the
accelerated particles onto the shocked fluid. 
The accelerated particles start to affect the fluid when their energy density
becomes comparable to the kinetic energy density of the fluid. In this
regime the test particle approximation breaks down, and the standard results 
of shock acceleration cannot be recovered. 
The only way to have a complete quantitative picture of this problem is
to use numerical simulations \cite{elli90,ebj95,ebj96,kj97}, but 
it is useful to have a practical analytical tool to understand the
physics behind the simulations and also to take into account the non linear
effects also when these simulations are not available, which is unfortunately
the rule rather than the exception.

Numerical simulations show that even when the fraction of particles
injected from the plasma is relatively small, the energy channelled into
these few particles can be close to the kinetic energy of the unshocked
fluid, making the test particle approach unsuitable. The most visible 
effect is on the spectrum of the accelerated particles, which shows
a peculiar hardening at the highest energies.

The need to have a theoretical understanding of the non-linear effects
in particle acceleration fueled many efforts in finding some
{\it effective} though simplified picture of the problem (see 
\cite{Blasi} for a discussion of these alternative approaches).
We will compare our findings with those of Ref. \cite{simple}.
%for instance
%\cite{dr_v80,dr_v81,dr_ax_su82,ax_l_mk82,ddv94} for a fluid apprach, 
%\cite{blandford80} for the case of small perturbations to the fluid,
%\cite{eich84a,eich84b,eich85,elleich85} for an approach somehow close to 
%the one presented here, \cite{simple} for a simple parametrization of the
%resulting distribution function, and finally \cite{malkov1,malkov2} for
%an alternative analytical solution).

In the present paper we summarize the results widely discussed in \cite{Blasi},
where an approach was proposed that provides a very nice fit to
the results of simulations and is in agreement with previous analyical 
calculations.

\section{Non Linear Shock Acceleration}

The distribution function of the particles accelerated at a planar infinite
shock can be written in implicit form as \cite{Blasi}:
$$
f_0(p)=\frac{N_{inj} q_s}{4\pi p_{inj}^3} \times
$$
\begin{equation}
\exp
\left\{-\int_{p_{inj}}^p \frac{dp}{p} \left[
\frac{3 u_p}{u_p-u_2} + \frac{1}{u_p-u_2}\frac{du_p}{d\ln p}
\right]\right\},
\label{eq:implicit}
\end{equation}
where we put $u_p=u_1 + \frac{1}{f_0(p)} \int_0^{\infty} dx
\left(\frac{du}{dx}\right) f(x,p)$, $q_s=\frac{3R_{sub}}{R_{sub}-1}$, and
$R_{sub}=u_1/u_2$ is the
compression factor at the shock surface [$u_1$ ($u_2$) is the fluid velocity 
upstream (downstream)].  $N_{inj}$ is the number density of
particles injected at the shock, parametrized here as $N_{inj}=\eta N_{gas,1}$,
where $N_{gas,1}$ is the gas density at $x=0^+$ (upstream). Particles are
assumed to be injected at the shock surface with momentum $p_{inj}$.
Eq. (\ref{eq:implicit}) tells 
us that the spectrum of accelerated particles has a local slope given by
$Q(p)=-\frac{3 u_p}{u_p-u_2} - \frac{1}{u_p-u_2}\frac{du_p}{d\ln p}$.
The problem of determining the spectrum of accelerated particles is then 
solved if the relation between $u_p$ and $p$ is found.

The thermodynamic properties of the shocked fluid are embedded in the usual
conservation equations, including now the contribution from accelerated
particles. 
The mass and momentum conservation equations read:
\begin{equation}
\rho_0 u_0 = \rho_p u_p,
\label{eq:mass}
\end{equation}
\begin{equation}
\rho_0 u_0^2 + P_{g,0} = \rho_p u_p^2 + P_{g,p} + P_{CR,p},
\label{eq:pressure}
\end{equation}
where $P_{g,0}$ and $P_{g,1}$ are the gas pressures at the point 
$x=+\infty$ and $x=x_p$ respectively, and $P_{CR,p}$ is the pressure
in accelerated particles at the point $x_p$ (we used the symbol $CR$
to mean {\it cosmic rays}, to be interpreted here in a very broad sense).
In writing eqs. (\ref{eq:mass}) and (\ref{eq:pressure}) we implicitly 
assumed that the average velocity $u_p$ coincides with the 
fluid velocity at the point where the particles with momentum $p$
turn around to recross the shock.
Our basic assumption is that
the diffusion is $p$-dependent and that therefore particles with 
larger momenta move farther away from the shock than lower momentum
particles. At each fixed $x_p$ only particles with momentum larger than $p$
are able to affect the fluid. 
Since only particles with momentum $> p$ can reach the point $x=x_p$, we can 
write $P_{CR,p} = \frac{4\pi}{3} \int_{p}^{p_{max}} dp p^3 v(p) f(p)$,
where $v(p)$ is the velocity of particles whose momentum is $p$, and 
$p_{max}$ is the maximum momentum achievable in the specific situation
under investigation. In realistic cases, $p_{max}$ is determined from 
geometry or from the duration of the shocked phase, or from the comparison
between the time scales of acceleration and losses. Here we consider it as a
parameter to be fixed {\it a priori}. From eq. (\ref{eq:pressure}) we
can see that there is a maximum distance, corresponding to 
the propagation of particles with momentum $p_{max}$ such that 
at larger distances the fluid is unaffected by the accelerated 
particles and $u_p=u_0$.
Assuming an adiabatic compression of the gas in the upstream region, 
we can write $P_{g,p}=P_{g,0} \left(\frac{\rho_p}{\rho_0}\right)^{\gamma_g}=
P_{g,0} \left(\frac{u_0}{u_p}\right)^{\gamma_g}$,
%\label{eq:Pgas}
%\end{equation}
where we used the conservation of mass, eq. (\ref{eq:mass}). The 
gas pressure far upstream is $P_{g,0}=\rho_0 u_0^2/(\gamma_g M_0^2)$,
where $\gamma_g$ is the ratio of specific heats ($\gamma_g=5/3$ for an
ideal gas) and $M_0$ is the fluid Mach number far upstream. 
Note that the adiabaticity condition cannot be applied at the shock jump, where
the adiabaticity condition is clearly violated.

After some algebra (se \cite{Blasi} for the details), the conservation 
equations imply the following equation:
$$
\ln {\cal D}U + \ln \left[1 - \frac{1}{M_0^2} U^{-(\gamma_g+1)}\right]
$$
$$
=\ln \left[ \frac{1}{3}\frac{N_{inj}q_s}{\rho_0 u_0^2} 
v(p) p_{inj}\right]+ 4\ln\left(\frac{p}{p_{inj}}\right) -
$$
\begin{equation}
\ln\left(\frac{R_{tot}U-1}{R_{sub}-1}\right)-
\int_{p_{inj}}^p \frac{dp}{p} \frac{3 R_{tot} U}{R_{tot} U - 1}, 
\label{eq:nobel}
\end{equation}
where $R_{tot}=u_0/u_2$, $U(p)=u_p/u_0$, and we put ${\cal D}U=dU/d\ln p$.
Solving this differential equation provides $U(p)$ and therefore the
spectrum of accelerated particles, through eq. (\ref{eq:implicit}).

The operative procedure for the calculation of the spectrum of accelerated
particles is simple: we fix the boundary condition at $p=p_{inj}$ such that 
$U(p_{inj})=u_1/u_0$ for some value of $u_1$ (fluid velocity at $x=0^+$). 
The evolution of $U$ as a function of $p$ is determined by 
eq. (\ref{eq:nobel}). 
The physical solution must have $U(p_{max})=1$ because at $p> p_{max}$
there are no accelerated particles to contribute any pressure. There is a 
unique value of $u_1$ for which the fluid velocity at the prescribed 
maximum momentum $p_{max}$ is $u_{p_{max}}=u_0$. Finding this value of
$u_1$ completely solves the problem, since eq. (\ref{eq:nobel}) provides
$U(p)$ and therefore the spectrum of accelerated particles.
Fig. 1 illustrates the distribution function (multiplied by $p^4$) for Mach 
number at infinity $M_0=43$, $p_{inj}=10^{-2} m c$ and $\eta=10^{-3}$ and
for $p_{max}=10^3 mc$ (solid line), $p_{max}=10^5 mc$
(dotted line) and $p_{max}=10^7 mc$ (dashed line), where $m$ is the mass
of the accelerated particles. The superimposed symbols 
show the corresponding results for the method in \cite{simple}.
\begin{figure}[htb]
%\vspace{9pt}
%\framebox[55mm]{\rule[-21mm]{0mm}{43mm}}
\includegraphics[width=8cm,height=6cm]{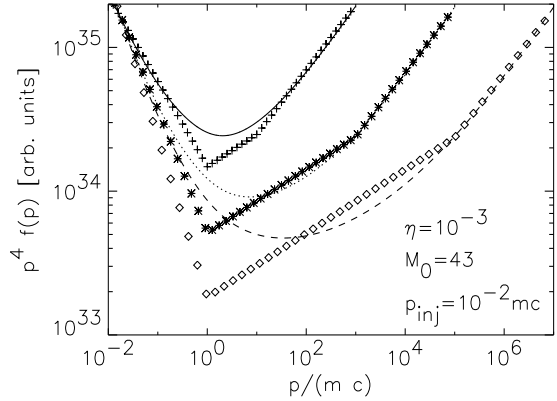}
\caption{Comparison between the prediction of our model (lines) and
those of Ref. \cite{simple} (symbols). The three sets of curves are obtained
for $p_{max}=10^3 mc$ (solid line and crosses), $p_{max}=10^5 mc$
(dotted line and stars) and $p_{max}=10^7 mc$ (dashed line and diamonds.}
\label{fig:Fig1}
\end{figure}
In Fig. 2 we plotted the results of our method for another set of parameters 
(indicated in the figure) and compared these results with the output of 
numerical simulations reported in \cite{simple}. It can be easily seen that the
agreement is impressive.
\begin{figure}[htb]
%\vspace{9pt}
%\framebox[55mm]{\rule[-21mm]{0mm}{43mm}}
\includegraphics[width=8cm,height=6cm]{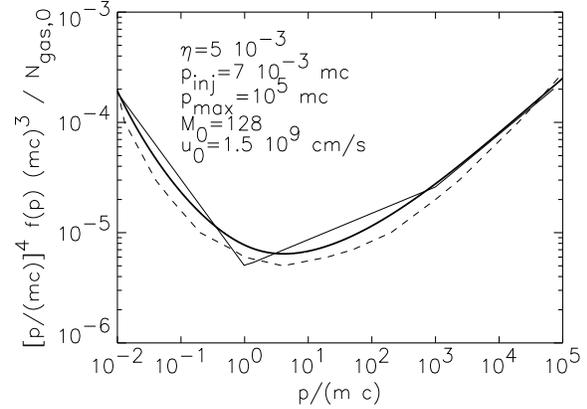}
\caption{Comparison between the predictions of our model (thick solid
line) and the results of simulations (dashed line) and the approximation in
Ref. \cite{simple} (solid light line).}
\label{fig:Fig2}
\end{figure}

\section{Conclusions}

We report on a novel semi-analytical 
approach to non linear shock acceleration, which improves some previous
attempts of other authors. This method is in good agreement with the 
previous approaches, and is also in impressive agreement with the results
of numerical simulations on shock acceleration. 
An extensive sets of predictions of this approach and a more complete
comparison with previous results are presented in \cite{Blasi}.
The most important phenomenological consequence of the inclusion of the
non linear effects in shock acceleration is the hardening of particle
spectra, which may reflect in a corresponding hardening of the spectra
of secondary particles (photons, electrons and neutrinos) generated in
the interactions of the accelerated particles with the environment.


\begin{thebibliography}{9}

\bibitem{be87} 
R.D. Blandford, and D. Eichler, Phys. Rep. 154 (1987) 1.

\bibitem{malkov} M.A. Malkov and L.O'C Drury, Rep. Progr. Phys. 64 (2001) 429.

\bibitem{elli90}
D.C. Ellison, E. M\"{o}bius, and G. Paschmann, Astrophys. J. 352 (1990) 376.

\bibitem{ebj95}
D.C. Ellison, M.G. Baring, and F.C. Jones, Astrophys. J. 453 (1995) 873.

\bibitem{ebj96}
D.C. Ellison, M.G. Baring, and F.C. Jones, Astrophys. J. 473 (1996) 1029.

\bibitem{kj97}
H. Kang, and T.W. Jones, Astrophys. J. 476 (1997) 875.

\bibitem{simple}
E.G. Berezhko, and D.C. Ellison, Astrophys. J. 526 (1999) 385.

\bibitem{Blasi}
P. Blasi, Astropart. Phys. {\it in press} [preprint astro-ph/0104064].

\end{thebibliography}
\end{document}